\documentclass[pre, reprint, 
 amsmath,amssymb,
 aps, superscriptaddress
]{revtex4-1}
\usepackage{graphicx}
\usepackage{amsmath}
\usepackage{amssymb}
\usepackage{enumerate}
\usepackage{color}
\usepackage[justification=raggedright,singlelinecheck=false,font=small]{caption}
\usepackage[justification=raggedright,singlelinecheck=false,font=small]{subcaption}

\def \be{\begin{equation}}
\def \ee{\end{equation}}
\def \ba{\begin{array}}
\def \ea{\end{array}}
\def \bea{\begin{eqnarray}}
\def \eea{\end{eqnarray}}

\newcommand{\bbeta}{{\mbox{\boldmath$\vec\beta$}}}
\newcommand{\bbetasub}{{\mbox{\boldmath$\vec{\scriptstyle\beta}$}}}
\newcommand{\Lbeta}{\ensuremath{\mathcal{L}_{\bbetasub}}}
\newcommand{\Lbetabar}{\ensuremath{\bar{\mathcal{L}}_{\bbetasub}}}

\begin{document}


\title{Information loss under coarse-graining: a geometric approach}

\date{\today}

\begin{abstract}
We use information geometry, in which the local distance between models measures their distinguishability from data, to quantify the flow of information under the renormalization group. We show that information about relevant parameters is preserved, with distances along relevant directions maintained under flow. By contrast, irrelevant parameters become less distinguishable under the flow, with distances along irrelevant directions contracting according to renormalization group exponents. We develop a covariant formalism to understand the contraction of the model manifold.  We then apply our tools to understand the emergence of the diffusion equation and more general statistical systems described by a free energy.  Our results give an information-theoretic justification of universality in terms of the flow of the model manifold under coarse-graining. 
\end{abstract}
\author{Archishman Raju}
\email{ar854@cornell.edu}
 \affiliation{Laboratory of Atomic and Solid State Physics, Cornell University, Ithaca, New York 14853}
\author{Benjamin B. Machta}
\email{benjamin.machta@yale.edu}
\affiliation{Department of Physics,  Yale University, New Haven CT 06511}
\affiliation{Systems Biology Institute,  Yale University, West Haven CT 06516}

\author{James P. Sethna}

\email{sethna@lassp.cornell.edu}
\affiliation{Laboratory of Atomic and Solid State Physics, Cornell University, Ithaca, New York 14853}
\date{\today}

\maketitle

Microscopically diverse systems often yield to surprisingly simple effective theories.  The renormalization group (RG) describes how system parameters change as the scale of observation grows and gives a precise explanation for this emergent simplicity.  On coarse-graining and rescaling, most parameter combinations are found to be \textit{irrelevant} and their importance decreases as the RG proceeds.  Effective theories are thus determined by a small number of \textit{relevant} parameters whose importance grows with increasing scale.
While the RG was initially used to understand systems with spatial symmetry, it has found applicability in a wide range of theories including avalanches~\cite{dahmen1996hysteresis}, turbulence~\cite{canet2016fully}, differential equations~\cite{chen1994renormalization}, the central limit theorem~\cite{jona1975renormalization, jona2001renormalization}, period doubling~\cite{feigenbaum1978quantitative} and quasi-periodic systems~\cite{rand1982universal, feigenbaum1982quasiperiodicity}. 

 Here we use information geometry~\cite{amari} to reformulate the RG as a statement about how the distinguishability of microscopic parameters depends on the scale of observation. The emergence of simple laws of the macroscopic behavior corresponds to a loss of information about the microscopics. A reformulation of the RG in the language of information geometry allows us to make a more precise statement about this correspondence. Relevant variables are distinguished as those for which there is no loss of information under coarse-graining. Hence, the underlying complexity is hidden because information about the irrelevant variables is lost under coarse-graining.   


Information geometry uses the Fisher Information Matrix (FIM) as a Riemannian metric on parameter space.  The resulting model manifold describes a space where local distance is an intrinsic measure of the distinguishability of nearby models (see ~\cite{mrugala1990statistical, ruppen2, quevedo2007geometrothermodynamics} for discussion of this and related thermodynamic metrics). The FIM is defined for a parameterized statistical model $P(\{x\}|\vec{\theta})$ giving the probability $P$ for a state $\{x\}$ and parameters $\theta^\mu$ as
\begin{align}
\label{eq:FIM}
g_{\mu\nu}&=-\sum_x P(x|\vec{\theta})\ \frac{\partial^2 \log(P(x|\vec{\theta}))}{\partial \theta^\mu\partial \theta^\nu}  , \\
&= -\sum_x P(x|\vec{\theta})\ \frac{\partial \log P(x|\vec{\theta})}{\partial \theta^\mu} \frac{\partial \log P(x|\vec{\theta})}{\partial \theta^\nu}
\end{align}
where $ds^2= g_{\mu\nu}\delta \theta^\mu \delta \theta^\nu$ defines the distinguishability of models that differ by $\delta \vec{\theta}$ from data $\{x\}$. 
The FIM is the natural metric for the space of probability distributions, in which our model manifold is embedded.

Our work builds on previous studies showing that in models from systems biology and elsewhere this metric has a characteristic \textit{sloppy} distribution, with eigenvalues spanning many orders of magnitude~\cite{marksloppy, markreview}.  It was recently shown that renormalizable models become sloppy, but only after their data, $\{x\}$ is coarsened~\cite{ben}, by decimation in the Ising model and blurring in time for diffusion.  While that paper was mostly numerical, we found that (1) `relevant' directions of the FIM do not grow, but instead are almost unaffected by coarsening and (2) irrelevant directions contract at a rate given by their RG exponent. 

In this letter, we address this question analytically and develop a covariant formalism to describe the flow of the model manifold. Past studies have considered the connection of the RG to geometry \cite{lassig1990geometry,o1993geometry, dolan1994geometrical}, but not to information theory. Here we use the RG to calculate the flow of the model manifold as the observation scale changes, and connect it to the distinguishability of parameters. We show that as coarse-graining proceeds relevant directions are exactly maintained while irrelevant directions contract.  Our results quantify the irreversibility of RG transformations: models which differ only in irrelevant directions rapidly become indistinguishable as the observation scale increases.  Our results also clarify that information about relevant directions is contained in large scale observables, since these directions do not contract as the observation scale increases. 
This striking feature, that relevant directions are preserved in the space of predictions, is distinct from the usual way the RG is thought about, where relevant directions grow in parameter space under coarse-graining.  

In Ref.~\cite{geompaper}, the connection between information geometry and the RG was discussed. The geometric meaning of scale invariance implies that the beta functions in the RG should be homothetic vectors.  Those authors calculated the form of the metric by assuming that distances are preserved under geometric flow determined by the beta functions.
As we show, this gives an essentially correct approximation to the metric only for relevant parameter combinations but this argument misses contributions to the metric whose form is not preserved under coarsening but which dominate the metric along irrelevant directions.   
This comes about because the scaling given in Ref.~\cite{geompaper} is only true for the singular part of the metric which diverges as the critical point is approached. An additional non-universal part, given by contributions to the metric which are analytic in the critical region, dominates along irrelevant parameter combinations. These contributions are central for the interpretation of coarse-graining that we offer below. In particular, information about irrelevant components is lost precisely because it is dominated by the analytic part of the metric.  

Our approach is inspired by information theory. Rather than deriving the form of the metric from the symmetries of the problem, we look at the evolution of the Fisher Information Metric under coarse-graining to see how information is lost. We first describe our formalism, apply it to the diffusion equation and then to coarse-graining classical statistical systems \footnote{We are not considering quantum systems where probabilistic interpretations are less clear \cite{vijay, beny2015information} } described by a free energy.


\begin{figure}[ht]
\begin {center}

		\includegraphics[width = 0.65\linewidth]{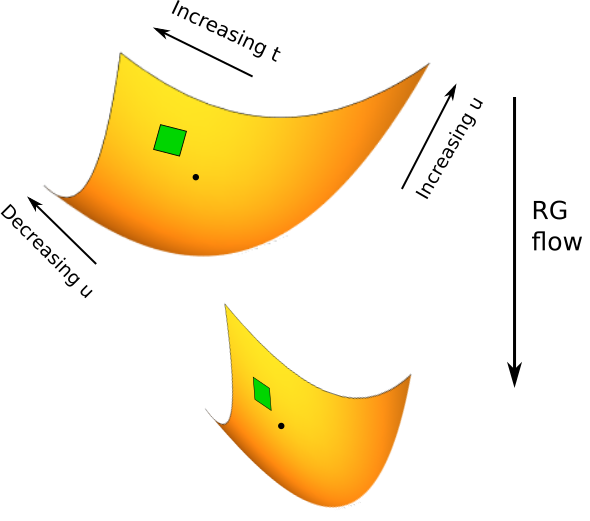}
	\caption{A cartoon figure of a section of the model manifold with one relevant ($t$) and one irrelevant ($u$) direction. The RG makes an initial patch (green) stay the same in the relevant direction but compress in the irrelevant directions. Information is preserved in relevant directions and lost in irrelevant directions. Finding an embedding to visualize the manifold is non-trivial \cite{quinn2017visualizing}.}
\label{invariantscurve}
\end {center}
\end{figure}

To understand how coarse-graining loses information about parameters, we consider an infinitesimal RG transformation of the form
 $d \theta^{\mu}/d b = \beta^{\mu}$
describing the flow of parameters by beta functions $\bbeta = \{\beta^\mu\}$
as the shortest length scale $b$ is increased. Sometimes, RG transformations are constructed to give exact Hamiltonians for macroscopic degrees of freedom. However, macroscopic measurements cannot infer all parameters that primarily influence the microscopics. It is this loss of information that we wish to quantify. Because information about microscopic degrees of freedom is discarded as the flow proceeds, there is no guarantee that a forward RG flow can be uniquely reversed.  However, the irreversibility of the RG is difficult to quantify through the RG equations alone.  
Instead, we turn to the metric tensor of eq.~\ref{eq:FIM}, a local measure of how hard it is to discriminate between models which differ by small changes of parameters~\cite{balasubramanian1997statistical}. We can quantify the loss of information in parameter space as microscopic information is discarded by asking how the metric tensor changes under an infinitesimal RG transformation.  
Since the metric is a two-tensor and the transformation is a flow given by a vector field, the answer is given by the Lie derivative $\Lbeta$ which defines the derivative of a covariant tensor carried along flow field $\bbeta$. 


Because the FIM quantifies the observability of parameters from a fixed amount of data, we consider an RG procedure which takes place on a large but finite system, whose physical size is fixed so that the observed system length $L$ shrinks during the RG according to $d L/d b = -L$.  As $L$ is not a parameter, we must add an additional term to the usual Lie derivative defining $\Lbetabar = \Lbeta - L \partial_L$. 
This additional term accounts for the reduction in the number of data
points because of the coarse-graining. The Fisher Information, through its definition in Eq 1, is linear in the amount of uncorrelated data. So coarse-graining by itself shrinks the metric tensor by $L \partial L$ as it
reduces the system size. (For system of dimension $d$, if $g = g_0 L^d$ is constant per unit
volume and has no RG flow, then $\partial g/\partial b = - L \partial L g = -L (d L^{d-1}) g_0 
= -d g$, so $g(b) = g_0 \exp(-d b) = g_0 L^d$.) This choice is more natural from our information theory perspective because coarse-graining cannot \textit{increase} information about any parameter combination. The change in the metric is then given by 
\begin{equation}
\label{liedef}
\Lbetabar g_{\mu \nu} = \beta^{\alpha} \partial_\alpha g_{\mu \nu} + g_{\alpha \mu} \partial_{\nu} \beta^{\alpha} + g_{\alpha \nu} \partial_{\mu} \beta^{\alpha} - L \partial_L g_{\mu \nu}.
\end{equation}
This modified Lie derivative $\Lbetabar$ is written as the sum of four terms. The first term is the directional derivative of the metric along the $\bbeta$ function, which keeps track of how the metric changes as the parameters flow. The next two terms come from the change in the parameter space distance between nearby points under flow. The last term describes the change in the metric due to the shrinking of the effective system size.  As long as the system is large enough to ignore the effects of the boundaries, the metric tensor $g_{\mu \nu} \propto L^d$, so that the first term can be written $-d g_{\mu \nu}$. Together, these terms describe how the metric changes under coarse-graining. Since we are working in a fixed parameterization, the change in the metric reflects a change in the proper distance between nearby systems and is an invariant quantity.  

To interpret this equation further, we consider the RG in its linearized form with eigenvalues $\lambda^{(\mu)}$, so that $\beta^{\mu} = \lambda^{(\mu)} \theta^{\mu}$ and we use the notation that the summation convention is not to be used if the indices are in parentheses. With $\partial_{\mu} \theta^{\alpha} = \delta^{\alpha}_{\mu}$, this yields a simplified equation
\begin{equation}
\label{lieequation}
 \Lbetabar g_{\mu \nu} =  \lambda^{(\alpha)} \theta^{\alpha} \partial_{\alpha} g_{\mu \nu} +  g_{\mu \nu} (\lambda^{(\mu)} + \lambda^{(\nu)}) - L \partial_L g_{\mu \nu} .
 \end{equation}
The RG is often done as a discrete operation, rather than a continuous one in which $\theta^\mu$ flow to $\theta^{\tilde \mu}$ in one discrete iteration.  In this case the new metric in the new parameters is given by $\tilde{g}_{\tilde \mu \tilde \nu} = \langle \partial_{\tilde{\nu}} \partial_{\tilde{\mu}} \log P (\tilde{x}) \rangle$. The usual goal of RG is to find an effective theory with renormalized parameters for macroscopic degrees of freedom. Our goal is to see how information about the original parameters transforms after coarse-graining, but we do sometimes find these renormalized parameters a useful intermediate, and when we use them we denote them with a $\sim$. To write the metric in bare parameters we simply change parameters according to $\tilde{g}_{\mu \nu} = \tilde{g}_{\tilde{\mu} \tilde{\nu}} \frac{\partial {\theta^{\tilde{\mu}}}}{\partial \theta^{\mu}} \frac{\partial {\theta}^{\tilde{\nu}}}{\partial \theta^{\nu}}$. Hence, the new metric is given by $\tilde{g}_{\nu \mu} = \langle \partial_{\tilde{\nu}} \partial_{\tilde{\mu}} \log P (\tilde{x}) \rangle \frac{\partial {\theta}^{\tilde{\mu}}}{\partial \theta^{\mu}} \frac{\partial {\theta}^{\tilde{\nu}}}{\partial \theta^{\nu}}$.

We illustrate this formalism by first considering the RG flow of a generalized hopping model of diffusion. Consider a particle undergoing stochastic motion according to a kernel $K_\tau(x)$ so that $$P(x(t+\tau)|x(t))= K_\tau\left(x(t+\tau)-x(t)\right).$$ We assume that $K_\tau$ has a finite second moment. How well can we infer the shape of $K_\tau$ by measuring time series $\vec{x}=\left\{x(0), x(\tau),x(2 \tau),...,x(m \tau) \right\}$? We can choose to parameterize the Kernel in different ways. A convenient way to parameterize probability distributions close to a Gaussian distribution is to use Hermite polynomials. Let $\theta^\mu$ parameterize an arbitrary Kernel $K_\tau$, and $H_\mu$ be the $\mu^{\textrm{th}}$ Hermite polynomial. For notational convenience, we define $\bar{H}_\mu = \frac{H_\mu(y)}{\mu!(\theta^2)^{\mu/2}}$. The Kernel is then given by 
\begin{equation}
K_{\tau}(x)=\frac{1}{\sqrt{2\pi (\theta^2)}} \exp{\left(-y^2/2\right)} \left[1 +\sum_{\mu \geq 3}  \theta^{\mu} \bar{H}_{\mu}(y) \right] , 
\label{kerneleq}
\end{equation}
where $y=(x-\theta^1) / (\theta^2)^{1/2}$. 

The probability of a time series $\vec{x}$ is given by 
$P(\vec{x})=\prod\limits_{i=1}^m K_\tau\left(x(i\tau)-x((i-1)\tau\right)$
and the FIM near $\theta^\mu = 0$ for $\mu>2$ is 
\begin{align}
g_{\mu\nu}^{(0)}
&=  \frac{m \delta_{\mu\nu}}{\mu! (\theta^2)^\mu}  , 
\end{align}
where the $(0)$ denotes a microscopic measurement. In typical measurements, the scale of observation is much larger than the natural timescale $\tau$.  How would the slowness of a measuring apparatus influence the ability to infer $K_\tau$?  One means to address this is to consider the renormalized kernel arising from the composition of $n$ time-steps. For example, after two steps
\begin{equation}
K_{2\tau}(x)=\int dx' K_\tau(x')K_\tau(x-x') .
\end{equation}

After $n$ steps, the kernel can be recast into the form of Equation~\ref{kerneleq} provided parameters are renormalized to $\theta^{\tilde \mu (n)}=n^{1-\mu/2} \theta^{\mu}$ where the spatial coordinate is rescaled according to $y^{(n)} = n^{-1/2} y$. The continuous limit here gives $d \theta^\mu/d \log n = (1 - \mu/2) \theta^\mu$ 
so that $\beta^\mu=(1 - \mu/2) \theta^\mu$.
The Kernel after $n$ steps becomes
\begin{equation}
K_{n\tau}(x)=\frac{1}{\sqrt{2\pi (\theta^2)}} e^{-\frac{(y^{(n)})^2}{2}} \left[1 +\sum_{\mu \geq 3}  n^{1-\frac{\mu}{2}} \theta^\mu \bar{H}_\mu(y^{(n)}) \right].
\end{equation}

We are interested in the case where observations occur at a scale much larger than $\tau$. In addition, for a fixed trajectory, the number of data points reduce by a factor of $n$, rescaling the metric in the discrete contribution corresponding to the third term in eq.~\ref{lieequation}. In the case of diffusion, the metric depends only on $\theta^2$ which is itself invariant under the RG operation. Hence, the first term in eq.~\ref{lieequation} is zero  so that the parameter dependence of the metric doesn't contribute to its change on rescaling. This is in part because of our choice of parameters.  In a different parameterization, this term could contribute to the change in the metric since the individual terms in eq.~\ref{lieequation} are not covariant. However, the Lie derivative as a whole is covariant so that the total change in the metric transforms covariantly under a change of parameters. Finally, the contribution of the second term in Eq.~\ref{lieequation} is given in a straightforward way with the eigenvalues $\lambda^{(\mu)} = (1 - \mu/2)$ and the new metric near the Gaussian fixed point the new metric is given by: 
\begin{align}
\tilde{g}_{\mu\nu}^{(n)} 
\label{diffusionmetric}
&= n^{1 - \mu} g_{\mu \nu}^{(0)}
\end{align}

For diffusion, the mean $\theta^{(1)}$ is the only relevant variable in the sense of having RG exponent $>0$, and its corresponding FIM eigenvalue is exactly preserved under this coarse-graining procedure. This makes sense:
the mean drift is given by the two endpoints, unchanged under coarse-graining.
Our main new prediction is that this is also true for the relevant variables
in general statistical mechanical systems.

The standard deviation is the next most distinguishable parameter, marginal in the RG sense, and under coarse-graining it becomes harder to see as quantified by the decreasing value of $g_{22}^{(n)}$ as $n$ increases. All other parameters become even more indistinguishable at late times. This framework quantifies the irreversibility of the RG through the inability to distinguish irrelevant variables at long time or length scales. 

Our approach emphasizes that the total information depends on the total amount of data. An alternative approach, more in line with statistical physics, would be to define a Fisher information density which is independent of the system size. Relevant directions would then spread out under coarse-graining with an exponent equal to the dimension of the system $d$, but \textit{not} to their RG exponent. This \textit{intensive} metric is discussed in Ref.~\cite{quinn2017visualizing}. Using the total information is not only more natural from an information-theoretic point of view but results (here and in the next section) correspond more to intuition: the information about relevant directions is preserved under coarse-graining.  

The diffusion equation is a simple example where everything can be analytically calculated. We now apply our formalism in more generality to statistical systems with a free energy.  
Consider a Boltzmann distribution for a micro state $x$, $P(x)=\exp (-H(x))/Z$, with Hamiltonian of the form $H = \sum_\mu \theta^\mu \Phi_\mu(x) $, where $\Phi_\mu(x)$ are functions of the micro state $x$ and where $\theta^\mu$ are parameters. Typically, $\Phi$s will be sums or integrals over space like $\int_\mathcal{V} \phi^2$ in Landau theory or $\sum_i x_i x_{i+1}$ in the Ising model. Because the Hamiltonian is in an exponential family and linear in its parameters the FIM can be written as (see eq.~\ref{eq:FIM}) \cite{ben}:
\begin{equation}
\label{FIMfreenergy}
 g_{\mu \nu} = \partial_{\mu} \partial_{\nu} \log \left(Z \right)= \langle \Phi_\mu \Phi_\nu \rangle -\langle \Phi_\mu \rangle \langle \Phi_\nu \rangle
\end{equation}
A renormalization group operation typically involves coarse-graining by tracing over some degrees of freedom (like in our previous example for diffusion). Let $\tilde{x}^{[n]}=C^{[n]}(x)$ be the coarse-grained state observed when the bare micro state is $x$. This coarse-graining could involve removing high momentum states, or decimating over alternate spins (see \cite{ben} for an example). For our purposes, it is only important that the coarse-graining acts as a map from bare micro-states to coarse-grained ones. We can define the restricted partition function $\tilde{Z}(\tilde{x}^{[n]}) =\sum_x \delta(C^{[n]}(x),\tilde{x}^{[n]}) e^{- H(x)}$ in terms of which the probability of being in coarse-grained state $\tilde{x}$ is $p(\tilde{x}) = \tilde{Z}(\tilde{x})/Z$. We choose our convention so that the renormalized Hamiltonian has the same form as the old one with renormalized parameters with any additional constant written separately
\begin{equation}
\label{hamiltonian}
\tilde{H}(\theta^{\tilde \mu}) - G(\theta^\mu) = H(\theta^\mu). 
\end{equation}
We can define the Fisher metric on the bare parameter space, but for coarse-grained observables $\tilde{x}$ using eq.~\ref{eq:FIM}: 
\begin{equation}
\label{eq:gb}
\tilde{g}^b_{\mu \nu} =  \partial_\mu \partial_\nu \log Z - \langle \partial_\mu \partial_\nu \log \tilde{Z}^b(x) \rangle 
\end{equation}
To calculate derivatives of $\log \tilde{Z}^b$ it is helpful to define the expectation value of an operator $\Phi(x)$ defined at the bare micro state, conditioned on coarse-graining to coarsened state $\tilde{x}^{[n]}$ as \cite{ben}
\begin{equation}
\left\{ \Phi(x) \right\}_{\tilde{x}^{[n]}}=\frac{1}{\tilde{Z}(\tilde{x}^{[n]})} \sum_x \delta(C^{[n]}(x),\tilde{x}^{[n]})\Phi(x) e^{- H(x)}
\end{equation}

We define correlation functions of these operators in the natural way, $\langle \{\Phi_1\}_{\tilde{x}^{[n]}} \{\Phi_2\}_{\tilde{x}^{[n]}} \rangle = \sum_{\tilde{x}^{[n]}} p(\tilde{x}^{[n]}) \{\Phi_1\}_{\tilde{x}^{[n]}} \{\Phi_2\}_{\tilde{x}^{[n]}} $. In this notation the reduction of the metric on coarsening is given by
\begin{equation}
\langle \partial_\mu \partial_\nu \log \tilde{Z}^b(x) \rangle = \langle \left\{ \Phi_\mu \Phi_\nu \right\}_{\tilde{x}^{[n]}} - \left\{ \Phi_\mu \right\}_{\tilde{x}^{[n]}}  \left\{ \Phi_\nu \right\}_{\tilde{x}^{[n]}} \rangle
\end{equation}
which is positive definite, demonstrating that coarsening reduces the FIM in all directions.  This form also contains intuition for our result derived below- that FIM components are almost preserved on coarsening for relevant directions, whose metric eigenvalues diverge near critical points. The divergence in these eigenvalues can be understood as arising due to the diverging range of correlations in the fields within the correlation functions in eq.~\ref{FIMfreenergy}. (each $\Phi$ is an integral over space.)  However,  the coarsened ensembles defined here are conditioned on fixing long wavelength modes or a subset of degrees of freedom, effectively screening correlations at lengths larger than the coarsening scale $b$.  As such correlations are absent at long scales so that $\langle \partial_\mu \partial_\nu \log \tilde{Z}^b \rangle$ must remain non-singular even as correlations in the uncoarsened fields diverge near the critical point.
The full coarsened FIM is given by:
\begin{equation}
\label{eq:gb2}
\tilde{g}^n_{\mu \nu} = \langle \left\{ \Phi_\mu \right\}_{\tilde{x}^{[n]}}  \left\{ \Phi_\nu \right\}_{\tilde{x}^{[n]}} \rangle - \langle \Phi_\mu \rangle \langle \Phi_\nu \rangle
\end{equation}

To see the origin of the loss of information, we can directly calculate $\tilde{g}_{\mu \nu} = \langle -\partial_\mu \partial_\nu \tilde{H} \rangle + \partial_\mu \partial_\nu Z$. Now, from Equations~\ref{hamiltonian} and ~\ref{FIMfreenergy}, we get $\tilde{g}_{\mu \nu} = g_{\mu \nu} - \partial_\mu \partial_\nu G$. The change in the Fisher Information Metric is a consequence of the analytic constant that gets added during the coarse-graining transformation. By exponentiating Equation~\ref{hamiltonian} and taking a trace, we can write this as $\exp(-F(\theta^\mu)) = \exp(G(\theta^\mu)) \exp(-F(\theta^{\tilde \mu}))$, where $F$ is the free energy. As is common, we find it useful to divide the total free energy into a singular and an analytic piece, $F = F^{(s)} + F^{(a)}$. Using our definition of the metric in Equation~\ref{eq:gb}, we can correspondingly write the metric $g_{\mu \nu} = g_{\mu \nu}^{(a)} + g_{\mu \nu}^{(s)}$. The singular part of the free energy is conserved along the flow. This can be written as 
\begin{equation}
\label{met}
  \beta^{\gamma} \partial_{\gamma} F^{(s)} = L \frac{\partial}{\partial L} F^{(s)} 
\end{equation}
  Taking two derivatives of this equation and assuming linearity of the beta functions, 
\begin{equation}
\label{sing1}
  \Lbetabar g^{(s)}_{\mu \nu} = 0 
\end{equation}
That is, metric components of the singular part of the metric are preserved along the flow. The relevant components of the metric are dominated by the singular part of the free energy and hence information about them is preserved along the flow. This is reflected in the divergence of quantities like the specific heat and susceptibility at the critical point. The irrelevant components of the metric are dominated by the analytic part of the free energy. To see this, consider a typical scaling form of the free energy of an Ising model with one irrelevant component $u$, $F = L^d t^{\frac{d}{\lambda_t}} \mathcal{F} (u t^{\frac{-\lambda_u}{\lambda_t}})$. If we assume $\mathcal{F}$ is analytic, we can take two derivatives to find the component of the metric $g^{(s)}_{u u} = L^d t^{\frac{d - 2 \lambda_u}{\lambda_t}} \mathcal{F}'' (u t^{\frac{-\lambda_u}{\lambda_t}})$. Since $\lambda_u$ is negative, this expression goes to $0$ at the critical point and is very small near to it. Away from the critical point, this has some finite value but is still small yielding well known corrections to scaling. 


We can calculate how the analytic part of the metric changes by going back to the transformation for the free energy $F^{(a)} (\theta^{\tilde \mu}) = F^{(a)} (\theta^\mu) + G(\theta^\mu)$. 
The change in the analytic part of the metric, after $n$ coarse-grainings, is given by $\Delta g_{\mu \nu}^{(a)}(\theta^{\tilde{\mu}(n)}) = - \sum_{i = 1}^{n-1} \partial_\mu \partial_\nu G(b^{m \lambda_{\alpha}} \theta^{\alpha})$. This can be simplified to 
\begin{equation}
 \Delta g_{\mu \nu}^{(a)}(\theta^{\tilde{\alpha}(n)}) = - \sum_{m = 1}^{n-1} b^{m (\lambda_{\mu} + \lambda_{\nu})} \partial_{\tilde \mu} \partial_{\tilde \nu} G(\theta^{\tilde \alpha})
\end{equation}
where  $\partial_{\tilde \mu} \partial_{\tilde \nu} G(\theta^{\tilde \alpha})$ is a positive definite quantity (as can be explicitly checked when calculation is possible, like in the 2-D Ising model on a triangular lattice \cite{2dising}). Hence, the analytic part of the free energy decreases with each coarse-graining. In this case, since the metric is analytic, the change in the metric because of the curvature (corresponding to the partial derivative in Equation~\ref{lieequation}) is less significant and the irrelevant components of the metric contract under the flow. 

One can write down a partial differential equation describing the flow of the singular part of the metric by the requirement that $\bar{\mathcal{L}} g^{(s)}_{\mu \nu} = 0$.  
\begin{equation}
\label{rgeq}
\beta^{\alpha} \partial_\alpha g^{(s)}_{\mu \nu} + g^{(s)}_{\alpha \mu} \partial_{\nu} \beta^{\alpha} + g^{(s)}_{\alpha \nu} \partial_{\mu} \beta^{\alpha} = d g^{(s)}_{\nu \mu}
\end{equation}

This equation determines the components of the metric in the relevant direction which can be obtained by solving this equation. Such equations have recently been solved for different kinds of RG flows near critical points in Ref~\cite{geompaper}. The change in the analytic part of the metric close to the critical point should be dominated by the change in parameter space distance in Equation~\ref{lieequation} (with higher order corrections). Together, this allows us to predict the scaling of the metric for linear beta functions with correlation length  $\xi$ and coarse-graining length $b$
\begin{align}
\label{sing2}
 g^{(s)}_{\mu \mu} &\sim L^d \xi^{-(d - 2 \lambda_\mu)} , \\
 \label{analytic2}
 g^{(a)}_{\mu \mu} &\sim L^d b^{-(d -  2 \lambda_\mu)} .
\end{align}
We note that this is a non-trivial result. It says that on ignoring some degrees of freedom in an Ising spin system (not observing them experimentally), one should be able to measure the temperature and magnetic field equally well from the coarsened system. The argument in statistical mechanics is somewhat technical but our confidence in the result comes from the numerical work in Ref.~\cite{ben} which foreshadowed the argument above.

To summarize, we have shown (eqs.~\ref{sing1} and~\ref{sing2}) that information about the relevant components, contained in the singular part of the metric, is preserved under RG flow whereas information about the irrelevant components, contained in the analytic part of the metric, reduces (eq.~\ref{analytic2}).
This is the information theoretic meaning of universality. The RG can be understood as a flow of the model manifold under which irrelevant components of theory becomes harder and harder to distinguish and can hence be ignored in a description of a theory at long length or time scales.


We note some assumptions that have gone into the analysis. We have assumed the $\bbeta$ functions that determine the RG flows are linear. In typical cases, it is possible to make analytic coordinate changes that make the $\bbeta$ functions linear. The Fisher information will then no longer be given by Equation~\ref{FIMfreenergy}, because the Hamiltonian will not be linear in the new variables.  It should be possible to incorporate these corrections. In other cases, the $\bbeta$ functions are inherently non-linear~\cite{raju2017renormalization} which leads to logarithmic and other corrections. It will be interesting to examine the significance of these inherent nonlinearities for information theory.

 Universal scaling behavior typically comes with analytic and singular \textit{corrections to scaling}. It is possible that such corrections can be organized by relevance using information theory by systematically including them in the free energy.  With the FIM as metric, the scalar curvature is small in the thermodynamic limit, scaling like $\xi^d/L^d$ with $\xi$ the correlation length, quantifying the intrinsic non-Gaussianity of fluctuations.  By using the Fisher Information density, the divergence of the Ricci curvature can be identified with the the divergence of the correlation length as occurs near critical points \cite{ruppen, ruppen2, geompaper2}. 
 
We expect that our information theoretic understanding of coarse-graining can help to identify relevant parameters of models even where explicit RG procedures are not available.  In Ref~\cite{mark}, unimportant directions were identified as thin directions of the model manifold and were systematically removed.  Here we find that relevant and irrelevant directions in parameter space behave differently under coarsening in a reparameterization invariant way, which can be seen even without an explicit parameter space flow. This suggests that relevant directions for more general coarsening procedures can be identified with FIM eigendirections which do not contract under coarsening.


AR and JPS were supported by the National Science Foundation through Grant No. NSF DMR-1312160 and DMR-1719490. BBM was supported by NSF~PHY~0957573 and a Lewis-Sigler Fellowship.  AR and BBM were supported in part by the International Centre for Theoretical Sciences (ICTS) during a visit for participating in the program US-India Advanced Studies Institute: Classical and Quantum Information (Code: ICTS/Prog-infoasi/2016/12).


\bibliographystyle {unsrt}
\bibliography{rginfo}



\end{document}